# Multiscale Governance

**David Pastor-Escuredo**[1,2,†] and **Philip Treleaven**[1]

[1]UCL, UK
[2]LifeD Lab, Madrid, Spain
[†]email: david@lifedlab.org

**ABSTRACT**

Future societal systems will be characterized by heterogeneous human behaviors and also collective action. The interaction between local systems and global systems will be complex. Humandemics will propagate because of the pathways that connect the different systems and several invariant behaviors and patterns that have emerged globally. On the contrary, infodemics of misinformation can be a risk as it has occurred in the COVID-19 pandemic. The emerging fragility or robustness of the system will depend on how this complex network of systems is governed. Future societal systems will not be only multiscale in terms of the social dimension, but also in the temporality. Necessary and proper prevention and response systems based on complexity, ethic and multi-scale governance will be required. Real-time response systems are the basis for resilience to be the foundation of robust societies. A top-down approach led by Governmental organs for managing humandemics is not sufficient and may be only effective if policies are very restrictive and their efficacy depends not only in the measures implemented but also on the dynamics of the policies and the population perception and compliance. This top-down approach is even weaker if there is not national and international coordination. Coordinating top-down agencies with bottom-up constructs will be the design principle. Multi-scale governance integrates decision-making processes with signaling, sensing and leadership mechanisms to drive thriving societal systems with real-time sensitivity.

## INTRODUCTION

The COVID-19 pandemic has shown the need of new governance schemas. Lessons from biological systems (Blanchard et al., 2009; Pastor-Escuredo et al., 2016) and digital epidemiology (Pastor-Escuredo, 2020; Salathe et al., 2012) can be applied to design new multi-scale governance. Tissue organization shows how coordination and signaling among cells leads to robust process of organogenesis (Keller, 2012; Keller, Davidson, & Shook, 2003). Network science is also a powerful framework used to characterize spreading patterns (Balcan et al., 2010; Boguná, Pastor-Satorras, & Vespignani, 2003; Meloni et al., 2011; Moreno & Vazquez, 2003; Pastor-Escuredo, 2020; Salathe et al., 2012). Integrating dynamics systems, networks and multi-scale systems it is possible to design new governance frameworks.

Humandemics are the different large-scale processes that propagate across the tangled network created through digital connectivity, world-wide economic and financial exchanges and multi-scale mobility. Managing, designing and leading governance constructs requires understanding and quantifying complexity as a previous step to exploit to create more resilient, robust and thriving societies. Multi-scale governance will have impact on different sectors and industries as a game-changing catalyzer. The changes in industry and sectors along with new boards tools will allow scalable transformations for sustainability. Demographics and urban-rural development will also play a key role on governance changing livelihoods (Zufiria et al., 2018), risk (Norman, Bar-Yam, & Taleb, 2020; Taleb, 2019), inequalities and trade.

This perspective presents how data-driven governance can be implemented as a multi-scale construct integrating local and global systems and also real-time alerts with long-term plan. We discuss several aspects that should be considered for such a framework and provide several recommendations based on learnings from social science and also developmental biology and biophysics that inspire the design of resilient systems (Pastor-Escuredo & del Alamo, 2020). In this light, we propose the integration of human-computation systems (David Pastor-Escuredo, 2021).

## FUTURE OF GOVERNANCE

Transformative processes require drivers and catalyzers to generate phase transitions and enable structural changes. These are the constitutive elements of multi-scale governance. Most current organizations and governance structures are stiff. Dynamic organizations must replace current organizations to be more sensitive and responsive to external phenomena, either positive or negative, to increase synergies and make good decisions.

Early warning mechanisms must be at the core of governance in private, public and social sectors. Although Artificial Intelligence and Data have been used as early warning (Kryvasheyeu et al., 2016; Pastor-Escuredo, Torres, Martínez-Torres, & Zufiria, 2020; Wilson et al., 2016) in humanitarian contexts, there is still a big gap to implement such systems in corporate governance and public management. Furthermore, besides alerts, governance requires simulations to understand effects and impact of actions to be taken or to design response in real-time but with clear response priorities. Current world challenged by biological, economic and environmental threats has to be progressively redesigned from the lens of dynamics and complexity. Dynamic organizations are more resilient and help build a more resilient and robust society and socio-economic tissue.

Coupled with alert systems, multiscale governance requires to set long-term objectives properly articulated with real-time response and decision making. Sustainability Development Goals are a proper framework to establish mid and long term objectives that are holistic and allow sectors to interact and find synergies to drive towards a more sustainable society. For these reasons, dynamic processes have to be led also by governments, multi-lateral actors and international institutions in collaboration with private sector corporations.

Regarding the spatial and human scope of governance, holistic frameworks powered by complex science used within multi-stakeholders ecosystems are the way forwards to create evidence-based governance that is effective for high socio-economic impact. Artificial Intelligence has been used to get insights based on data but normally in a closed environment and with limited socio-economic scope. Private sector governance will likely keep this trend as privacy and competitive interest hamper sharing data at scale.

Federated learning (Yang, Liu, Chen, & Tong, 2019) is a potential solution to manage privacy and competitive interests through decentralized access to data and algorithmic agents. However, centralization of indicators is also required to generate transversal insights that are necessary for policy makers, public leaders, new innovators and entrepreneurs and different types of social and humanitarian agencies. The proper solution will rely on different levels of utility-privacy where data can be queried depending on its potential positive impact, mitigating privacy risk and augmenting social benefits (Pastor-Escuredo et al., 2020)

Data sharing and federated learning are key as future governance requires a transition to generate wide insights that open perspectives and encourage collaboration and transformative projects. Machine Learning, data science and visual analytics should evolve to create this knowledge ecosystems that are sound scientifically but also have high social value. Depth and wideness are two elements to promote in decision making processes through learning and training of decision makers and leaders.

Evidence-based policy making and AI-driven systems require also ethical frameworks to be acceptable and useful. An example is the use of digital apps for contact tracing that have risen social and ethical concerns that depend on the technological design, their deployment, their governance and their application (Altmann et al., 2020; Berman, Carter, Herranz, & Sekara, 2020; Vinuesa, Theodorou, Battaglini, & Dignum, 2020). Human-centered design is required, but also a broader perspective of an acceptable digitalization both in terms of technology and technology governance. Typical technical values such as transparency, efficiency and trustworthiness are necessary but not sufficient to be the ground for evidence-based policy making in the digital era. Frameworks will depend, of course, on cultural particularities, but it is important to have global frameworks for governance for a more coordinated society that can achieve sustainability goals.

Beyond evidence-based decision making, digital technology can help building more decentralized, responsive, flexible and accountable systems for governance. Although current debate points towards the bias of AI, probably, the most promising use of digital technology is to help building better public and private organizations to overcome human limitations (Pastor-Escuredo & Vinuesa, 2020). Algorithmic governance can mediate in power relationships and open spaces for collective intelligence if properly designed and deployed, rather than making more obscure and less interpretable decision making systems (David Pastor-Escuredo, 2021).

**INTERACTIONS BETWEEN LAYERS OF GOVERNANCE**

Only through more dynamic governance it will be possible to build up resilience as a necessary requirement for sustainability by implementing policies, mobilizing investment, proper regulations, international agreements, etc. Resilience and robustness require parts of the system to activate ahead and react to external conditions facilitating effective response and driving necessary and constructive transformations and change. We may think of such mechanisms as genes that activate and are specialized to response to specific stimuli.

Governance involves several actors starting from workers to governance boards, shareholders, stakeholders, investors and regulators and policy makers. These actors are organized in different layers that interact with each other in complex ways. An initial approach would be to model these actors as agents to simulate governance. However, technology should be oriented to change how these different actors interact. Current power structures, even when legitimate, do not offer guarantees of such kind of response due to structural stiffness, overload of hierarchies and bureaucracy, uncertainty for decision making, wicked power relationships and responsibility ownership.

Technology can help make better connections between the different layers of governance, from pro-active participation of citizens, clients and workers to tools for boards and investors to make better decisions or react better to emerging problems. The key aspect of multiscale governance is the data that can be collected integrating both objective outcomes, incomes and results with subjective data about actors' perception such as staff morale, clients' perception or competition environment.

Data-driven governance has to fill the gap between internal audits with ecosystem understanding. Quantifying internal corporate structure and dynamics interlinked with external phenomena is key for boards and decision makers. As discussed, data-intensive applications are risky in terms of privacy and competitive interests. Algorithmic governance can help by quantifying the level of utility and risk of data to be used for decision making within corporations and across corporations and sectors. Through data it is possible to understand the corporate internal structure and the socio-economic tissue and eventually model how corporate-level decisions would affect the societal tissue.

In this regard, multilayered networks (Aleta & Moreno, 2019; De Domenico, Solé-Ribalta, Omodei, Gómez, & Arenas, 2015) are a proper tool to model interactions intra and inter organizations. Data is the missing element to populate these models. Having structured data of organizations interactions would allow computation to better model how sectors can cooperate, how organizations can implement structural changes, how economy can be better stimulated, etc. Federated learning will allow to progress towards interconnected socio-economic networks for data-driven governance.

Crises and pandemics have also highlighted that in a hyper connected world, there is a severe lack of international coordination systems even within the European Union. Alliances and partnerships are key for new sectors and activities to emerge and also create opportunities to more types of innovation and entrepreneurship.

## COMMUNITY OF PRACTICE

New governance is not possible without new profiles of workers, directors and catalyzers. The integration of scientists on governance platforms organically is a requirement to harness a secure future of data-driven governance. Beyond expert committees, science and technology must penetrate in the organizations and decision-making processes. Thus, technology in governance is bidirectional: scientists would need training on governance and non-academic incentives and directors must embrace new technology for their activity.

Furthermore, algorithmic governance enabled by Blockchain can totally change the way boards of directors operate. Decentralized Autonomous Organizations are a model of new governance that may become a trend in larger corporations in different sectors. Although there is much debate about ethics and impact frameworks of algorithms, algorithms have a good chance to make more transparent decisions. Algorithmic governance, of course, will require also monitoring systems to ensure positive impact. The governance of algorithmic governance is itself an emerging paradigm that will gather much of the intellectual efforts in the years to come (Jobin, Ienca, & Vayena, 2019; Rahwan et al., 2019; Theodorou & Dignum, 2019).

Interdisciplinary, international and independent (and even decentralized) teams should help make decisions, interpret results and analyze outcomes. In that regard, COVID-19 has been a promising milestone of academic collaboration (Luengo-Oroz et al., 2020) that is necessary to leverage the necessary knowledge, technology and resources in a timely and effective manner. Collaboration will be necessarily layered integrating different actors as a network that needs management where digital technology should help as well.

AI and data could help activate the network through signaling processes and governance automation as it has started happening in the humanitarian sector (Pastor-Escuredo et al., 2020). A certain level of automation in partnerships and collaboration including data sharing would relieve from responsibility burdens of decision makers to accelerate processes. This is rather controversial with some AI regulation experts, but responsibility should never lead to inaction and digitalization is an opportunity to make not only more transparent but also more actionable and committed governance, always under the supervision of ethics and accountable impact.

Finally, citizens will be also core in future governance. Through collective action and decision platforms (Lévy & Bononno, 1997; Malone, 2018; Mulgan, 2018), citizens will be able to have influence in laws, investment, evaluation and planning. Through sensing and communication, collective response and action can be implemented. Governance complexity must be addressed through new structures and organizations designed through the lens of complexity (Pastor-Escuredo & Tarazona, 2021).

Computation and technology should be the basis of governance frameworks that leverage complexity, innovation and decentralization. Technology is key at several steps. First, as part of ideation process, computation can stimulate originality, innovation and entrepreneurship to generate better and more innovative ideas. Contrast of ideas with real-world problems is a good way to make sure that talent is at the service of deep transformations. Team building is another important element where computation can help creating robust teams within organizations and across organizations to ensure innovation, transparency, accountability and peoples' fulfillment. Ideas also need filtering to synthesize which are the most appropriate and suitable for being implemented. Computation and data-driven systems are a good way to select ideas and establish priorities. Idea selection is specially relevant in a world with resources that need to be properly managed towards sustainability. Impact prediction and evaluation are the other modules that can be also based on computation to manage teams, ideas and resources and enable learning and evolutive improvement.

# CONCLUSION

Multiscale governance implies thinking in organizations and places as part of complex systems where they interact with each other giving rise to collective actions. Collectiveness is not contradictory with organization, however, new multi-scale organization systems are required to manage the complexity and state of emergency of the world we live in. Leadership and networks are fundamental to build a more responsive and robust socio-economic tissue. Leadership is required to drive change, but we need to consider two fundamental aspects in leadership: ideas and dynamics of the networks. Thus, several organizations can play leading roles at certain times in an orchestrated complex system that aims at maximizing social benefit through the activation of actors as a genetic program.

This type of framework is not only good for existing structures and organizations, but it is also necessary for innovation and drive constructive efforts in the private sector, specially in ecosystems of start-ups. A new type of socio-economic tissue based on collaboration that aligns collective efforts in specific challenges and missions with a proper scientific and quantitative frameworks (Mazzucato, Kattel, & Ryan-Collins, 2020). From biological systems we can learn that a certain level of specialization combined with interactions and sensing mechanisms are the foundation for emergent properties such as sustainability.

Science based policy and governance is and emergent opportunity to develop more sustainable, resilient, robust and ethical societies. However, a simplified version of data-driven and evidence-based mechanisms can lead to very negative outcomes and also generate distrust in the population against technology and computation-driven decisions. Multi-disciplinary research and innovation is required to develop data-driven governance at all scales.

When considering complex and systemic problems such as pandemics, holistic approaches should prevail over linear decision making. The design of computation-powered mechanisms should promote both depth and wideness of perspectives to improve judgement and wisdom of population, teams and boards rather than instrumental tools to keep doing "business as usual". Current AI approaches may provide certain depth in data analysis, but they lack the wideness, which may lead into decisions with uncontrolled impact. Novel computational frameworks are required to manage complexity and public-private interactions.

The human side of governance and policy making cannot be overlooked. Collaborative efforts have to be integrated into governance platforms with capacities and capabilities to propose deep transformations and new mechanisms that are required for response to systemic threats. Technology should be the basis for activation, data sharing, collaboration, exchanges and also mediate in partnerships to catalyze responsible action. Intra-organization and inter-organization relationships management is a unique opportunity for a world that is interconnected and has computational frameworks to help decision.

Another key issue is the lack of leadership, both in terms of action and also generating narratives. The unbearable responsibility for many decision makers and politicians of causing harm (socio-economic, cultural, spiritual and health) through strict policies and disruptive measures is likely causing inaction in many local and national governments. We must open the chance for computational governance to complement existing mechanisms and allow disruption at scale.

Collaboration and coordination should be the ground for governance. However, these platforms need to introduce disruptive elements based on organizational and technological innovation. Necessarily, this process must be undertaken laterally and vertically considering the local problems of people but incrementing the awareness of the systemic problems we face. Minorities can lead to unfair decisions if there is a lack of proper integrative governance and ethical frameworks. Sustainability goals must be achieved in an inclusive way, so acceleration and inclusiveness must be coupled into the actions for new policies and governance platforms. This is not only an ethical principle, but a design, implementation and deployment principle for an interconnected world that is resilient and sustainable. Ethical principles that promote protection, action and future projection should be the basis of evidence-based systems to not become empty computational boxes.

Designing a societal system based on multiscale governance will be a process that requires experimentation and space for failure demanding also robustness of the system to mitigate negative outcomes and impacts.